\documentclass[reprint,amsmath,amssymb,aps,showpacs]{revtex4-1}

\usepackage{texdraw}
\usepackage{graphicx}
\usepackage{dcolumn}
\usepackage{bm}
\usepackage{subfigure}
\usepackage{url}
\usepackage{multirow}

\begin{document}


\title{Re-examining Archie's law: Conductance description by tortuosity and constriction}

\author{Carl Fredrik Berg}
\address{Statoil R\&D Center, Arkitekt Ebbels veg 10, Rotvoll, 7053 Trondheim, Norway}
\email{cfbe@statoil.com}

\begin{abstract}
In this article we investigate the electrical conductance of an insulating porous medium (e.g., a sedimentary rock) filled with an electrolyte (e.g., brine), usually described using the Archie cementation exponent. We show how the electrical conductance depends on changes in the drift velocity and the length of the electric field lines, in addition to the porosity and the conductance of the electrolyte. We characterized the length of the electric field lines by a tortuosity and the changes in drift velocity by a constriction factor. Both the tortuosity and the constriction factor are descriptors of the pore microstructure. We define a conductance reduction factor to measure the local contributions of the pore microstructure to the global conductance. It is shown that the global conductance reduction factor is the product of the tortuosity squared divided by the constriction factor, thereby proving that the combined effect of tortuosity and constriction, in addition to the porosity and conductance of the electrolyte, fully describes the effective electrical conductance of a porous medium. We show that our tortuosity, constriction factor, and conductance reduction factor reproduce the electrical conductance for idealized porous media. They are also applied to Bentheimer sandstone, where we describe a microstructure-related correlation between porosity and conductivity using both the global conductance reduction factor and the distinct contributions from tortuosity and constriction. Overall, this work shows how the empirical Archie cementation exponent can be substituted by more descriptive, physical parameters, either by the global conductance reduction factor or by tortuosity and constriction.
\end{abstract}

\pacs{47.56.+r,91.60.Tn}

\maketitle

\section{Introduction}\label{sec:introduction}

The electrical conductivity of a brine filled sedimentary rock is an essential part of electrical resistivity log interpretation in petroleum exploration \cite{helander1983fundamentals}. Its importance has led to a substantial amount of research, without resulting in any satisfactory theoretical description of the connection between the pore microstructure and the electrical conductivity of the rock 
.

The electrical conductivity of a porous medium (e.g., a sedimentary rock) consisting of an insulating matrix and an electrolyte (e.g., brine) with constant conductivity depends upon the electrical conductivity of the fluid, relative volumes of the matrix and the fluid (porosity), and upon the geometrical distributions of the matrix and the fluid (pore microstructure) relative to the applied electrical potential \cite{dullien1992porous}. The \emph{formation resistivity factor} $F$ is defined to be the ratio of the electrical resistivity of the rock saturated with brine $R_o$ to the resistivity of the brine $R_w$: 
$$F=\frac{R_o}{R_w}.$$ 
In \cite{archie1942electrical} Archie formulated an empirical relationship between the formation resistivity factor $F$ and the porosity $\phi$ to describe the conductivity of a porous medium. This relationship was given by the equation $F=\phi^{-m}$, where $m$ is the \emph{cementation exponent}.

We will consider the electrical conductivity of porous media filled with an electrolyte, or likewise the conductance of networks of resistors. A network might represent a simplification of the pore microstructure of a porous medium, where pore bodies corresponds to the nodes of the network and pore throats to the resistors \cite{fatt1956network}. Network analogues has been widely used to represent natural porous media such as reservoir rocks \cite{blunt2001flow,oren1998extending,bauer2011computed}.

Several definitions for tortuosity exist. A common definition of the tortuosity $\tau$ for a path is the ratio between the distance $l$ between the endpoints of the path and the path length $s$, i.e.\ $\tau=l/s$ (see e.g., \cite{adler1992porous}). It is a measure of deviation from the shortest possible path. For an idealized porous media consisting of a single circular tube of constant cross-sectional area and length $s$ connecting the opposite sides of a cube of side length $\Delta S$ we have $F = 1/(\tau^2 \phi)$, where $\phi$ is the porosity and $\tau = \Delta S/s$ is the tortuosity of the tube \cite{dullien1992porous}.

In an electrolyte the mobile charged particles move randomly, however the drift (average) velocity $\vec{v}$ is given by the equation $\vec{v} = \mu \mathbf{E}$ for an electro static field $\mathbf{E}$, where $\mu$ is the electrical mobility. Since $\mathbf{E} = -\nabla \Phi$ for the electrical potential $\Phi$, we have $\vec{v} = -\mu \nabla \Phi$. For example, for a straight circular tube the drift velocity $v$ is proportional to $I/A$, where $I$ is the current and $A$ the cross-sectional area. Varying the cross-sectional area gives a varying drift velocity, hence energy is expended and the conductivity of the tube is reduced. Assume we have a tube of length $l$ and cross-sectional area $A(x)$ at point $x$ that is filled with an electrolyte with constant conductivity $\sigma$. The conductance of the tube is then $G_r = \sigma/\int_0^l 1/A(x) dx$. A tube with equal volume and equal length, but with a constant cross-sectional area, would have a cross-sectional area $\overline{A}= \int_0^l A(x) dx / l$ and a conductance $G_o = \sigma \overline{A} / l = \sigma \int_0^l A(x) dx / l^2$. The inverse of the reduced conductivity due to variation in the cross-sectional area is given by the \emph{constriction factor} \cite{owen1952resistivity,boyack1963theory} 
\begin{equation}
C = G_o/G_r = \frac{1}{l^2} \int_0^l A(x) dx \int_0^l \frac{1}{A(x)} dx.
\label{eq:const_old_def}
\end{equation}
For an idealized porous medium consisting of a single straight circular tube with varying cross-sectional area, we then have $F = C/ \phi$.

Since the current is constant in the pore channel, we have $v(x) \propto I/A(x)$, so 
\begin{align}
C &= \frac{1}{l^2} \int_0^l A(x) dx \int_0^l \frac{1}{A(x)} dx \notag \\
 &= \frac{1}{l^2} \int_0^l v(x) dx \int_0^l \frac{1}{v(x)} dx \notag \\
&= \frac{1}{l^2} \int_0^l \lVert \nabla \Phi(x) \rVert dx \int_0^l \frac{1}{ \lVert \nabla \Phi(x) \rVert} dx.
\label{eq:constriction_def}
\end{align}
In the following we use Eq. \eqref{eq:constriction_def} as our definition of the constriction factor, which is then a reformulation of the more common definition given by Eq. \eqref{eq:const_old_def} \cite{owen1952resistivity,boyack1963theory,dullien1992porous}.  This definition also holds when the cross-sectional area is ill defined. The constriction factor is a measure of changes in drift velocity.

Note that variation in cross-sectional area also introduces tortuosity. For simplicity we disregarded the tortuosity introduced by constrictions when we related the constriction factor to the formation factor above. Likewise, tortuosity might introduce changes in drift velocity; for simplicity this was disregarded when we related tortuosity to the formation factor above.

The aim of this article is to investigate the physical relation between the conductance of a porous medium filled with an electrolyte and its microstructure. In Sec. \ref{section:crf} we introduce the concept of the conductance reduction factor of a porous medium or a network of resistors, which is used to describe the local contribution to the global conductance. In Sec. \ref{section:tort_const} we show how this global conductance can be described by distinct contributions of tortuosity and constriction, both inherent to the microstructure of the porous medium. The relation among the global conductance reduction factor, tortuosity, and the constriction factor is demonstrated through idealized porous media in Sec. \ref{section:idealized_por}. In Sec. \ref{section:bentheimer_network} we use Bentheimer sandstone data to demonstrate these relations also for natural porous media. 

\section{Conductance reduction factor derivation}\label{section:crf}

Consider a cubed porous medium $V$ of side length $\Delta s$ consisting of an insulating matrix (without surface conductance) and a pore space $\Omega \subset V$ filled with an electrolyte with constant conductance $\sigma$, and a potential drop $\Delta \Phi$ over two opposite sides of the cube. We define the \emph{conductance reduction factor} as 
\begin{equation}
\iota =  \lVert \nabla \Phi \rVert \frac{\Delta s}{\Delta \Phi}.
\label{eq:iota_def}
\end{equation}

As mentioned, the potential gradient gives rise to an average drift velocity of the mobile charged particles. The drift velocity is a conservative vector field, and the \emph{electric field lines} are paths that are everywhere parallel to the drift velocity (equivalently, the electric field). Consider an electric field line of length $\Delta s$ between inlet and outlet and with constant drift velocity. Then $\lVert \nabla \Phi \rVert = \Delta \Phi / \Delta s$, so $\iota = 1$ on the given electric field line. In the following we refer to a pore geometry giving a constant conductance reduction factor of $\iota = 1$ as \emph{optimal (conducting) geometry}. A set of straight circular tubes oriented in the direction of the applied potential and with a constant cross-sectional area gives an optimal geometry. An optimal geometry has maximum conductance for the given pore volume (porosity), and the conductance is $1/R_o= \phi/R_w$, so 
\begin{equation}
F = \frac{R_o }{R_w} = \frac{1}{\phi}.
\label{eq:optimal_frf}
\end{equation}

For more complex pore geometries the factor $\iota$ is no longer constant in general. We define the \emph{global conductance reduction factor} $\iota^2_g$ as the volume weighted average of the local conductance reduction factor $\iota^2$:
\begin{equation}
\iota^2_g = \frac{1}{\Omega} \int_\Omega \iota^2 d\Omega.
\label{eq:global_iota_def}
\end{equation}
In the following we show that for any porous medium we have $F = 1/(\iota^2_g \phi) \geq 1/\phi$. Hence $\iota^2_g \phi = \phi^m$, where $m$ is the Archie cementation exponent. Note that the conductance reduction factor $\iota$ might be larger than $1$, while $\iota^2_g \leq 1$. 

Following an electric field line from inlet to outlet, the potential drop along the field line equals the total potential drop over the sample; hence the potential can be treated as a measure of the distance between inlet and outlet. Comparing the distance on a electric field line ($ds$) for a given potential drop ($d\Phi$) to the distance in an optimal geometry given the same potential drop gives the fraction
$$ \frac{d\Phi}{ds}\frac{\Delta s}{\Delta \Phi}.$$
The limit $ds \to 0$ then yields Eq. \eqref{eq:iota_def}. We will see later that the conductance reduction factor measures the combined effect of tortuosity and changes in drift velocity on the conductivity of the porous medium.

We next investigate the conductance reduction factor for porous media and then turn our attention to network analogs.

\subsection{Porous media}

\begin{figure}
\includegraphics[width=7.649cm]{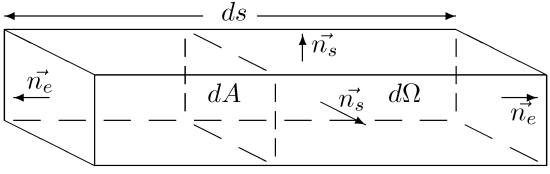}
\caption{An infinitesimal volume element $d\Omega$ where $\vec{n_s} \cdot \nabla \Phi = 0$ and $\vec{n_e} \cdot \nabla \Phi = \pm \lVert \nabla \Phi \rVert$.}
\label{box_image}
\end{figure}

For a cubed insulating porous medium of side length $\Delta s$ filled with an electrolyte with constant conductivity $\sigma$, let a potential drop $\Delta \Phi$ be applied over two opposite side planes of the cube. For a point $x$ inside the pore space $\Omega$, assume an infinitesimal volume element $d\Omega$ around $x$ as in Fig. \ref{box_image}.  We will orient $d\Omega$ so that the normal vector for the side planes $\vec{n_s}$ is orthogonal to the gradient of the potential drop, $\vec{n_s} \cdot \nabla \Phi = 0$, hence the volume is oriented so that there is no current over the side planes. Also the normal vector for the end planes $\vec{n_e}$ is in the same or opposite direction as the gradient of the potential drop, $\vec{n_e} \cdot \nabla \Phi = \pm \lVert \nabla \Phi \rVert$. The potential drop between the end planes of $d\Omega$ will be denoted $d\Phi$. When $d\Omega \to 0$ the lengths of the electric field lines contained in $d\Omega$ converges, and their length will be denoted $ds$. Also the cross-sectional areas in $d\Omega$ will converge when $d\Omega \to 0$, and will be denoted by $dA$. The resistance over $d\Omega$ is then given by $dR = ds/(\sigma dA)$.

We will now investigate the corresponding volume $d \Omega'$ in an optimal geometry that gives equal current $dI' = dI$ given the same potential drop $d\Phi' = d\Phi$, where all marked parameters are from the optimal geometry. Invoking Ohm's law gives $dR' = d\Phi'/dI' =  d\Phi/dI = dR$. Since $\iota' = 1$ in the optimal geometry, we have $ds' = \Delta s d\Phi'/\Delta \Phi  = \Delta s d\Phi /\Delta \Phi $. Using 
$$\frac{ds'}{\sigma dA'} = dR' = dR = \frac{ds}{\sigma dA},$$
we then obtain
\begin{align*}
dA' ds' &= \frac{ds'dA}{ds} ds' = \frac{dA}{ds}\left(d\Phi \frac{\Delta s}{\Delta \Phi} \right)^2 \\
&= \left(\frac{d\Phi}{ds}\right)^2 \left(\frac{\Delta s}{\Delta \Phi}\right)^2 dA ds.
\end{align*}
Since $dA'ds' = d\Omega'$, we obtain
\begin{align*}
\lim_{d\Omega \to 0} \frac{d\Omega'}{d\Omega} &= \lim_{d\Omega \to 0}  \left(\frac{d\Phi}{ds} \frac{\Delta s}{\Delta \Phi}\right)^2 \\&=  \left(\lVert \nabla \Phi (x) \rVert \frac{\Delta s}{\Delta \Phi}\right)^2 = \iota^2.
\end{align*}
This leads to
\begin{equation}
\Omega' = \int_\Omega d\Omega' = \int_\Omega \iota^2 d\Omega = \iota^2_g \Omega,
\end{equation}
where $\Omega'$ is the volume needed in an optimal geometry to obtain the same electrical conductance as the porous medium with pore space $\Omega$. Invoking Eq. \eqref{eq:optimal_frf} this yields 
\begin{equation}
F = \frac{V}{\Omega'} = \frac{V}{\iota^2_g \Omega} = \frac{1}{\iota^2_g \phi},
\label{eq:frf_iota_glob}
\end{equation}
where $V$ is the total volume of the porous medium.
Hence we have derived a relationship between the local conductance reduction factor and the global conductance reduction factor that describes the global conductance.

We might assume that for every point $x \in \Omega$ we have continuous paths in $\Omega$ to the two sides where the potential drop is applied. All parts of the pore space not satisfying this condition can be assumed to be part of the matrix, since it will not contribute to the conductance. However, since $\iota^2(x)=0$ for all points $x \in \Omega$ not satisfying the condition above, the relation $F = 1/(\iota^2_g \phi)$ still holds when such points are included in $\Omega$.

\subsection{Network model}
Consider a network representation of a porous medium consisting of resistors (pore throats) and nodes (pore bodies) connecting the resistors \cite{fatt1956network}. A resistor $t$ has an associated length $s(t)$ and a constant cross section $A(t)$. The nodes do not have associated volumes, they are only connecting the resistors. In such a network consider a resistor $t$ of volume $V(t)=A(t)s(t)$, and with potential drop over its length $\Delta \Phi(t)$. The resistance over $t$ is then $R(t)=s(t)/(\sigma A(t))$. For the whole sample we denote the potential drop $\Delta \Phi$ and the size of the sample in the direction of the global potential drop $\Delta s$.

We now investigate the corresponding volume $V'(t)$ in an optimal geometry that has equal current $\Delta \Phi(t)/R(t)$ given the same potential drop $\Delta \Phi(t)$. Then $R'(t) = R(t)$, which leads to 
$$\frac{s'(t)}{\sigma A'(t)} = R'(t) = R(t) = \frac{s(t)}{\sigma A(t)}.$$
Using that $\iota'=1$ in the optimal geometry, we have $s'(t) = \Delta \Phi(t) \Delta s /\Delta \Phi$. Then, similarly to the porous medium, we find
\begin{align*}
A'(t) s'(t) &= \frac{s'(t)A(t)}{s(t)} s'(t) = \frac{A(t)}{s(t)}\left(\Delta \Phi(t) \frac{\Delta s}{\Delta \Phi} \right)^2 \\
&= \left(\frac{\Delta \Phi(t)}{s(t)}\right)^2 \left(\frac{\Delta s}{\Delta \Phi}\right)^2 A(t) s(t) = \iota(t)^2 A(t) s(t).
\end{align*}
Hence $V'(t) = A'(t) s'(t) = \iota(t)^2 A(t) s(t) = \iota(t)^2 V(t)$ is the volume needed in optimal geometry to obtain equal current given the same potential drop as in resistor $t$.

Let $\Omega = \Sigma V(t)$ and $\Omega' = \Sigma V'(t)$ be the total network volume and the volume needed in an optimal geometry to transport the same current, respectively. For a network we define 
\begin{equation}
\iota^2_g = \frac{1}{\Omega} \sum_t{ V(t) \iota(t)^2} = \frac{1}{\Omega} \sum_t{ V'(t)} = \frac{\Omega'}{\Omega}.
\end{equation}
Hence the global conductance reduction factor $\iota^2_g$ for the whole sample is a volume weighted average of the local conductance reduction factors $\iota(t)^2$. With $\Omega' = \iota^2_g \Omega$ and invoking Eq.\ \eqref{eq:optimal_frf}, we have 
\begin{equation}
F = \frac{V}{\Omega'} = \frac{V}{\iota^2_g \Omega} = \frac{1}{\iota^2_g \phi}.
\end{equation}
where $V$ is the total volume of the porous medium. 

It is important to note that the local components of the global conductance reduction factor are inter-related since the local potential drop is dependent on the full network. Nevertheless, the conductance reduction factor offers the possibility to investigate the local contributions to the global conductance.

\section{Tortuosity and constriction}\label{section:tort_const}

It is well established that tortuosity and constriction affect the conductance of a porous medium \cite{helander1983fundamentals,dullien1992porous,winsauer1952resistivity}. What we show next is \emph{how} tortuosity and constriction affect conductance.

As before, consider a cubed insulating porous medium of side length $\Delta s$ filled with an electrolyte with constant conductance and an applied potential drop $\Delta \Phi$ over two opposite sides of the cube. We define the tortuosity $\tau$ of this porous medium as the side length of the cube divided by the average electric field line length. This is given by
\begin{equation}
\tau = \frac{\Delta s}{\int_\Omega \frac{dI}{I_t} ds} = \frac{1}{\frac{1}{I_t}\int_\Omega \frac{\sigma \lVert \nabla \Phi (x) \rVert}{\Delta s} d\Omega},
\label{eq:tort}
\end{equation}
where $\Omega$ is the pore space and $d\Omega$ is an infinitesimal volume element oriented in alignment with the drift velocity (see Fig. \ref{box_image}), $dI$ is the current transported through $d\Omega$, and $I_t$ is the total current through the porous medium. Here we use that 
$$\lim_{d\Omega \to 0} \frac{dI ds}{d\Omega} = \lim_{d\Omega \to 0} \sigma \frac{d\Phi}{ds} = \sigma \lVert \nabla \Phi \rVert.$$
Note that $ds$ is the length of the infinitesimal $d\Omega$ in the direction of the drift velocity, so the integral in Eq.\ \eqref{eq:tort} yields a current-weighted average of lengths in the direction of the drift velocity. In other words, the integral gives the average distance the current has to traverse.

Since we introduce a measure for constriction along electric field lines, we need a measure for tortuosity along electric field lines too. Let $\Gamma \in \mathbb{P}$ be an electric field line, where $\mathbb{P}$ is the set of all electric field lines, let $l_\Gamma$ be the length of the electric field line $\Gamma$, and let $\Omega_c = \{ x \in \Omega \vert x \in \Gamma \in \mathbb{P} \} = \{ x \in \Omega \vert \nabla \Phi(x) \not=0 \}$ be the subset of $\Omega$ given by the set of electric field lines or, equivalently, the conducting subset where $\nabla \Phi \not=0$. We define the conducting porosity as the fraction $\phi_c = \Omega_c / V$. The tortuosity $\tau_c$ of the porous medium is given by
\begin{equation}
\tau^2_c = \frac{1}{\Omega_c}\int_\mathbb{P} \tau(\Gamma)^2 d\Gamma = \frac{1}{\Omega_c}\int_\mathbb{P} \left( \frac{\Delta s}{l_\Gamma} \right)^2 d\Gamma.
\label{eq:tort_path}
\end{equation}
Here $\tau(\Gamma)=\Delta s/l_\Gamma$ is the tortuosity of the electric field line $\Gamma$. We can, without loss of generality, assume that the set $d\Gamma$ is a simply connected space (i.e., there are no holes in $d\Gamma$). This assumption is used throughout this article.

The constriction factor was introduced in Sec. \ref{sec:introduction} and measures changes in drift velocity. The constriction factor $C(\Gamma)$ for an electric field line $\Gamma$ of length $l_\Gamma$ is given by the integral
\begin{align}
C(\Gamma) &= \frac{1}{l_\Gamma^2} \int_\Gamma \lVert \nabla \Phi(s) \rVert ds \int_\Gamma \frac{1}{ \lVert \nabla \Phi(s) \rVert} ds \notag \\
&= \frac{1}{l_\Gamma^2} \Delta \Phi \int_\Gamma \frac{1}{ \lVert \nabla \Phi(s) \rVert} ds,
\label{eq:const_one_path}
\end{align}
where we use that $\lim_{ds\to 0} d\Phi/ ds = \lVert \nabla \Phi \rVert$ on the electric field line $\Gamma$. We define the constriction factor $C_c$ as the current-weighted average of the electric field line constriction factors $C(\Gamma)$:
\begin{equation}
C_c = \frac{1}{I_t} \int_\mathbb{P} C(\Gamma)dI_\Gamma.
\label{eq:const}
\end{equation}
Here $dI_\Gamma$ is the infinitesimal current to the infinitesimal set of electric filed lines $d\Gamma$, and as before $I_t$ is the total current through the porous medium.

We then obtain $\iota_c$ by restricting $\iota_g$ as given by Eq. \eqref{eq:global_iota_def} to $\Omega_c$:
\begin{align}
\iota^2_c &= \frac{1}{\Omega_c} \int_{\Omega_c} \iota^2 d\Omega = \frac{1}{\Omega_c} \int_\mathbb{P} \iota(\Gamma)^2 d\Gamma \notag \\
&= \frac{1}{\Omega_c} \int_\mathbb{P} \frac{1}{d\Gamma} \int_{d\Gamma} \iota^2 d\Omega d\Gamma.
\label{eq_iota_vs_iota_path}
\end{align}
Here $\iota(\Gamma)^2 = \lim_{d\Gamma \to \Gamma} 1/d\Gamma \int_{d\Gamma} \iota^2 d\Omega$ is the conductance reduction factor of the electric field line $\Gamma$.
Note that 
\begin{equation}
\iota^2_c = \iota^2_g\frac{\phi}{\phi_c} = \iota^2_g\frac{\Omega}{\Omega_c}
\label{eq:iota_conducting}
\end{equation}
since $\iota(x) = 0$ for $x \in \Omega \setminus \Omega_c$.

By choosing all $d\Omega$ such that they contain all electric field lines in $d\Gamma$, we have the following correspondence:
\begin{align}
\iota(\Gamma)^2 &= \lim_{d\Gamma \to \Gamma} \frac{1}{d\Gamma} \int_{d\Gamma} \iota^2 d\Omega \notag \\
&= \lim_{d\Gamma \to \Gamma} \frac{1}{\int_\Gamma dA(s) ds} \int_{\Gamma} \left( \frac{d\Phi}{ds} \right)^2 \left( \frac{\Delta s}{\Delta \Phi} \right)^2 dA(s) ds \notag \\
&= \left( \frac{\Delta s}{\Delta \Phi} \right)^2 \frac{1}{\int_\Gamma \frac{ds}{d\Phi} ds} \int_{\Gamma} \left( \frac{d\Phi}{ds} \right)^2  \frac{ds}{d\Phi} ds \notag \\
&= \frac{1}{\frac{1}{l_\Gamma^2} \Delta \Phi \int_\Gamma \frac{1}{\lVert \nabla \Phi \rVert} ds} \left( \frac{\Delta s}{l_\Gamma} \right)^2 = \frac{\tau(\Gamma)^2}{C(\Gamma).} \label{eq:iota_eq_tortconst_one_path}
\end{align}
Here we use the correspondence $dI_\Gamma = \sigma dA(s) d\Phi / ds$, where as before $dI_\Gamma$ is the infinitesimal current to the infinitesimal set of electric field lines $d\Gamma$. Also, as before the set $d\Gamma$ is simply connected. The cross-sectional area of $d\Gamma$ at point $s$ along the electric field line $\Gamma$ is denoted $dA(s)$.

We now have the framework needed for our main result; combining Eqs. \eqref{eq:tort_path}, \eqref{eq:const}, and \eqref{eq:iota_eq_tortconst_one_path} we obtain
\begin{equation}
\frac{\tau^2_c}{C_c} = \frac{\frac{1}{\Omega_c}\int_\mathbb{P} \tau(\Gamma)^2 d\Gamma}{\frac{1}{I_t} \int_\mathbb{P} C(\Gamma)dI_\Gamma} = \iota_c^2 \frac{\int_\mathbb{P} \tau(\Gamma)^2 d\Gamma}{\int_\mathbb{P} C(\Gamma) \iota(\Gamma)^2 d\Gamma} = \iota_c^2.
\label{eq:iota_eq_tortconst_global}
\end{equation}
Here we employ the equality $$I_t = \frac{\Delta A \Delta \Phi \iota^2_c \phi_c \sigma}{\Delta s} = \frac{\Omega_c \Delta A \Delta \Phi \iota_c^2 \sigma}{V \Delta s}$$
and the corresponding equality
$$dI_\Gamma = \frac{d\Gamma \Delta A \Delta \Phi \iota(\Gamma)^2 \sigma}{V \Delta s.}$$

For a porous medium, combining Eqs \eqref{eq:frf_iota_glob}, \eqref{eq:iota_conducting}, and \eqref{eq:iota_eq_tortconst_global}, we then arrive at the equality 
\begin{equation}
F = \frac{1}{\iota^2_g \phi} = \frac{1}{\iota^2_c \phi_c} = \frac{C_c}{\tau^2_c \phi_c}.
\label{eq:iota_tort_const_eq}
\end{equation}
The constriction factor $C_c$ measures the loss of conductance due to changes in drift velocity, and the tortuosity $\tau^2_c$ measures the loss of conductance due to electric field line lengths. Given by the equality above, these two factors fully describe the electrical conductance for a given porosity $\phi_c$ and conductance of the electrolyte $\sigma$. The tortuosity and constriction factor are describing the pore microstructure relative to the applied electrical potential, thus they give a theoretical description of the connection between the pore microstructure and the electrical conductivity of the porous medium.

If we assume that for every point $x \in \Omega$ we have continuous paths in $\Omega$ to the two sides where the potential drop is applied, then $\nabla \Phi(x) \not=0$ almost everywhere in $\Omega$, so the closure of the set $\Omega_c$ equals $\Omega$. However, solutions for the electric potential in a grid representation of the porous medium might still give $\nabla \Phi =0$. Also, in resistor network analogs where every element has network paths both to the inlet and to the outlet, we might have $\nabla \Phi =0$, for example, when nodes are connected to exactly one resistor. Hence $\Omega_c \subseteq \Omega$ in general. However, when $\phi = \phi_c$, we have
\begin{equation}
F = \frac{1}{\iota^2_g \phi} = \frac{C_c}{\tau^2_c \phi}.
\end{equation}
We perform calculations of tortuosity $\tau_c$ and constriction factor $C_c$ on resistor network analogs of porous media in Sec. \ref{section:bentheimer_network}.

\section{Examples from idealized porous media}\label{section:idealized_por}

Let us return to the idealized porous media described in Sec. \ref{sec:introduction}, consisting of one circular pore channel connecting to opposite side planes of a cube of side-length $\Delta S$, and with a potential drop $\Delta \Phi$ over the side planes connected by the pore channel.
For all examples in this section we then have $\Omega_c = \Omega$, $\phi_c = \phi$, $\iota_c = \iota_g$, $C_c = C$ and $\tau_c = \tau$. Therefore we do not use the $c$ subscripts.

\subsection{Conductance reduction factor and tortuosity}

\begin{figure}
\includegraphics[width=6cm]{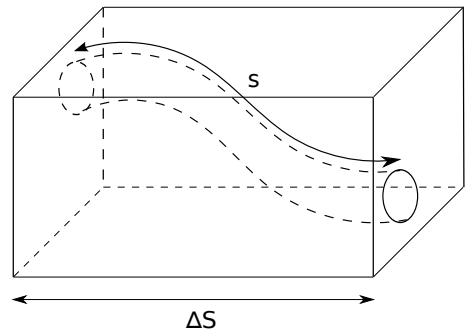}
\caption{An idealized porous medium consisting of one tube of length $s$ with constant cross-sectional area inside a cube of side-length $\Delta S$.}
\label{tort_tube}
\end{figure}

Consider a tube of length $s$ and constant cross-sectional area, as depicted in Fig. \ref{tort_tube}. The magnitude of the gradient of the potential inside the pore channel is constant at $\Delta \Phi / s$, thus
$$\iota^2 = \lVert \nabla \Phi \rVert^2 \left(\frac{\Delta S}{\Delta \Phi}\right)^2 = \left(\frac{\Delta \Phi}{s}\right)^2 \left(\frac{\Delta S}{\Delta \Phi}\right)^2 = \left(\frac{\Delta S}{s}\right)^2 = \tau^2.$$
Since $\iota$ is constant throughout the pore channel, we obtain 
\begin{equation}
F = \frac{1}{\iota^2_g \phi} = \frac{1}{\iota^2 \phi} = \frac{1}{\tau^2 \phi}.
\end{equation}

\subsection{Conductance reduction factor and constriction}

In this subsection, we consider an idealized model of constriction. Consider one straight circular pore channel in the direction of the potential drop, hence of length $\Delta S$, with cross-sectional area $A(s)$ for $s\in [0,\Delta S]$. In the following we neglect the tortuosity introduced in such a medium due to the constriction, as mentioned in Section \ref{sec:introduction}. Since the electrical current $I = I(s) = \sigma A(s) \frac{d\Phi}{ds} = \sigma A(s) \lVert \nabla \Phi (s) \rVert$ is constant, we get 
$$\Delta \Phi = \int_0^{\Delta S} d\Phi = \frac{I}{\sigma} \int_0^{\Delta S} \frac{ds}{A(s)},$$ $$\lVert \nabla \Phi(s) \rVert =\frac{I}{\sigma A(s)}=\frac{\Delta \Phi}{A(s)\int_0^{\Delta S} 1/A(s) ds}.$$
 Then the conductance reduction factor is given by
\begin{align*}
\iota(s)^2 &= \lVert \nabla \Phi (s)\rVert^2 \left(\frac{\Delta S}{\Delta \Phi}\right)^2 \\
 &= \left(\frac{\Delta S}{A(s)\int_0^{\Delta S} 1/A(s) ds}\right)^2.
\end{align*}
By volume averaging the conductance reduction factor $\iota(s)^2$, we have
\begin{align*}
\iota^2_g &= \frac{1}{\Omega} \int_0^{\Delta S} \iota(s)^2 A(s) ds \\
&= \frac{1}{\int_0^{\Delta S} A(s) ds} \int_0^{\Delta S} \frac{\Delta S^2}{\left(\int_0^{\Delta S} \frac{1}{A(s)} ds\right)^2 \left(A(s)\right)^2} A(s) ds \\
&= \frac{\Delta S^2}{\int_0^{\Delta S} A(s) ds \int_0^{\Delta S} \frac{1}{A(s)} ds} = 1/C.
\end{align*}
This yields 
\begin{equation}
F = \frac{1}{\iota^2_g \phi} = \frac{C}{\phi}.
\end{equation}

\subsection{Conductance reduction factor, tortuosity, and constriction}\label{subsect:iota_tort_const}

Let us turn to a combination of tortuosity and constriction. We now consider a single circular pore channel with cross-sectional area $A(s)$ for $s\in [0,s]$, where $s$ is the length of the tube. In line with the calculations in the previous subsection, substituting the length of the channel $\Delta S$ with the new length $s$, we obtain
$$\iota(s)^2 = \left(\frac{\Delta S}{A(s)\int_0^s 1/A(s) ds}\right)^2,$$
\begin{align*}
\iota^2_g &= \frac{1}{\int_0^s A(s) ds} \int_0^s \frac{\Delta S^2}{(\int_0^s\frac{1}{A(s)} ds)^2 (A(s))^2} A(s) ds \\
&= \left(\frac{\Delta S}{s}\right)^2\frac{s^2}{\int_0^s A(s) ds \int_0^s \frac{1}{A(s)} ds} = \tau^2/C,
\end{align*}
thus yielding 
\begin{equation}
F = \frac{1}{\iota^2_g \phi} = \frac{C}{\tau^2 \phi}.
\end{equation}

\section{Bentheimer rock example}\label{section:bentheimer_network}

In this section we use a micro-CT (microtomography) image and rock models of a Bentheimer sandstone to compute the global conductance reduction factor $\iota^2_g$, the tortuosity $\tau_c$, and the constriction factor $C_c$. Using the commercially available software e-Core \cite{e-core_v133} we generated three-dimensional rock models of a Bentheimer sandstone with porosities ranging from 5\% to 25\%. The software simulates the processes of grain sedimentation and diagenesis \cite{oren2006digital}. The same grain sedimentation model parameters were used for all models. We achieved different porosities by changing the amount of quartz cementation \cite{pilotti2000reconstruction}, as occurring in diagenetic rock forming processes, while other processes like compaction could also naturally lead to a porosity reduction. The model sample size was 2.5 mm$^3$ with a resolution of 5 $\mu$m, compared to the micro-CT image of 3.7x3.7x2.4 mm with a resolution of 3.7 $\mu$m.

\begin{figure}
\includegraphics[width=8cm]{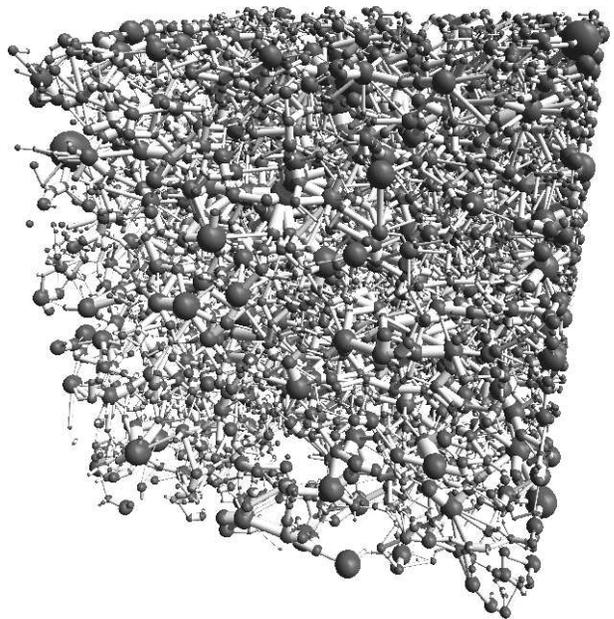}
\caption{A visualization of the network representation of the pore space of Bentheimer rock model \emph{f} with 15.4\% porosity. Balls represent pore bodies, and sticks represent pore throats. In this network there are 9260 pore bodies and 14792 pore throats.}
\label{network_img}
\end{figure}

Network representations of the pore microstructure were numerically extracted from both the generated models and the micro-CT image. The algorithm segments the grid representation of the pore microstructure into pore bodies and pore throats, each associated with a volume and a point $x \in V$. In Fig. \ref{network_img} we have visualized one such network. The distance $\lVert x_{t1} - x_{t2} \rVert$ between the two pore bodies $t1$ and $t2$ connected by a pore throat $t3$ is divided into parts: parts $l_{t1}$ and $l_{t2}$ associated with each pore body, $t1$ and $t2$, respectively; and part $l_{t3}$ associated with the pore throat, where $l_{t1} + l_{t3} + l_{t2} = \lVert x_{t1} - x_{t2} \rVert$. We associate tubes with a constant cross-sectional area with these parts: For the middle tube the cross-sectional area is $V_{t3}/l_{t3}$ with length $l_{t3}$, while for the tubes associated with the pore bodies $t1$ and $t2$ the cross-sectional areas are $V_{ti}/(\alpha_{ti} l_{ti})$ with lengths $l_{ti}$, where $V_{ti}$ is the volume of the pore body or throat as given by the segmentation algorithm, and $\alpha_{ti}$ is the number of pore throats connected to pore body $ti$. These tubes then have conductances $G_{it} = \sigma V_{ti}/(l_{ti}^2 \alpha_{ti})$ for $i=t1,t2$ and $G_{t3} = \sigma V_{t3}/l_{t3}^2$.

\begin{table*}
\begin{center}
\caption{Model and micro-CT results.}
\label{tab_models}
\begin{tabular}{|c|c|c|c|c|c|c|c|c|c|}
\hline
&\multicolumn{2}{c|}{Porosity}&\multicolumn{2}{c|}{Conduction reduction factor}&\multicolumn{2}{c|}{Tortuosity}&Constriction &Formation resistivity &Cementation  \\
Rock&$\phi$&$\phi_c$&$\iota^2_g$&$\iota_c^2$&$\tau$&$\tau_c$&$C_c$&factor $F$&exponent $m$\\
\hline
a	&	0.049	&	0.028	&	0.017	&	0.030	&	0.430	&	0.399	&	5.290	&	1186.03	&	2.355	\\
b	&	0.076	&	0.053	&	0.038	&	0.054	&	0.471	&	0.441	&	3.574	&	347.18	&	2.272	\\
c	&	0.096	&	0.077	&	0.083	&	0.103	&	0.542	&	0.519	&	2.604	&	125.80	&	2.064	\\
d	&	0.122	&	0.106	&	0.116	&	0.133	&	0.569	&	0.546	&	2.241	&	71.02	&	2.024	\\
e	&	0.138	&	0.124	&	0.135	&	0.151	&	0.588	&	0.567	&	2.126	&	53.43	&	2.011	\\
f	&	0.154	&	0.139	&	0.150	&	0.165	&	0.595	&	0.576	&	2.011	&	43.51	&	2.014	\\
g	&	0.159	&	0.146	&	0.153	&	0.167	&	0.602	&	0.581	&	2.028	&	41.03	&	2.020	\\
h	&	0.177	&	0.166	&	0.172	&	0.184	&	0.609	&	0.589	&	1.888	&	32.83	&	2.016	\\
i	&	0.199	&	0.188	&	0.193	&	0.204	&	0.632	&	0.611	&	1.832	&	26.04	&	2.020	\\
j	&	0.221	&	0.211	&	0.203	&	0.212	&	0.633	&	0.612	&	1.769	&	22.33	&	2.056	\\
k	&	0.241	&	0.233	&	0.217	&	0.225	&	0.645	&	0.624	&	1.731	&	19.07	&	2.073	\\
l	&	0.254	&	0.247	&	0.219	&	0.226	&	0.646	&	0.624	&	1.726	&	17.97	&	2.109	\\
Micro-CT	&	0.193	&	0.169	&	0.171	&	0.195	&	0.629	&	0.612	&	1.924	&	30.40	&	2.075	\\
\hline
\end{tabular}
\end{center}
\end{table*}

We next transform the network model to a resistor network analog. There is a one-to-one correspondence between the pore throats in the porous medium and the resistors in the resistor network analog; also, there is a one-to-one correspondence between the pore bodies and the network nodes. Each resistor $t$ is given a conductance 
\begin{align}
G_t & = \left(\frac{1}{G_{1t}}+\frac{1}{G_{3t}}+\frac{1}{G_{2t}}\right)^{-1} \notag \\
&= \sigma\left(\frac{l_{t1}^2 \alpha_{t1}}{V_{t1}}+\frac{l_{t3}^2}{V_{t3}}+\frac{l_{t2}^2 \alpha_{t2}}{V_{t2}}\right)^{-1},
\end{align}
where subscripts $t1$ and $t2$ represent the pore bodies connected to throat $t3$. In contrast with e.g., Eq. (18) in Ref. \cite{valvatne2004predictive}, this conductance formulation ensures that the resistor network analog has the same pore volume as the corresponding rock model or micro-CT data.

Let $\Phi_i$ be the potential corresponding to node $i$, and $t_{ij}$ the resistors connected to node $i$. We then solve for $\Phi_i$ such that 
\begin{equation}
\sum_{j=1}^{\alpha_i}{ G_{t_{ij}}(\Phi_i-\Phi_j)} = 0,
\end{equation}
where inlet and outlet nodes have fixed potential as boundary conditions. This is equivalent to Kirchhoff's circuit laws for the conservation of charge and energy in electrical circuits. The solution technique is equivalent to solving a resistor network problem.

The resistor $t$ is viewed as consisting of the three tubes $t1, t3$, and $t2$ with a constant cross-sectional area in series. We find potentials $\Phi_{it}$ for $i=1,2$ such that $G_{t1} \lvert \Phi_{t1}-\Phi_{1t} \rvert = G_{t3} \lvert \Phi_{1t}-\Phi_{2t} \rvert = G_{t2} \lvert \Phi_{2t} -\Phi_{t2} \rvert$. This enables the calculation of the conductance reduction factor for the three tube sections: 
$$\iota^2_{ti} = \frac{\left(\frac{\Phi_{ti}-\Phi_{it}}{l_{ti}}\right)^2}{\left(\frac{\Delta \Phi}{\Delta s}\right)^2} \text{ for } i=1,2 \text{ and } \iota^2_{t3} = \frac{\left(\frac{\Phi_{1t}-\Phi_{2t}}{l_{t3}}\right)^2}{\left(\frac{\Delta \Phi}{\Delta s}\right)^2}.$$
The volume average of these local contributions then gives the global conductance reduction factor,
\begin{equation}
\iota_g^2 = \frac{1}{\Omega} \sum_t{ \left( \frac{V_{t1}}{\alpha_{t1}} \iota^2_{t1} + \frac{V_{t2}}{\alpha_{t2}} \iota^2_{t2} + V_{t3} \iota^2_{t3} \right) }.
\end{equation}
The calculated results are reported in Table \ref{tab_models}.

\begin{figure}
\includegraphics[width=8cm]{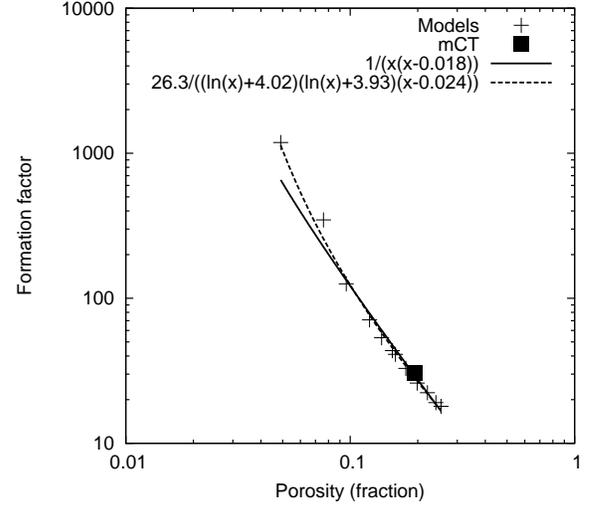}
\caption{Plot showing porosity $\phi$ versus formation resistivity factor $F$ for both rock models and micro-CT (mCT) data, together with two functions describing the correlation.}
\label{por_vs_frf}
\end{figure}

In Table \ref{tab_models} we have also listed the formation resistivity factor as $F=1/(\iota^2_g \phi)$. Fig. \ref{por_vs_frf} is a plot of the porosity and formation resistivity factor for the network representations. Here the micro-CT data fall in line with the rock models.

\begin{figure}
\includegraphics[width=8cm]{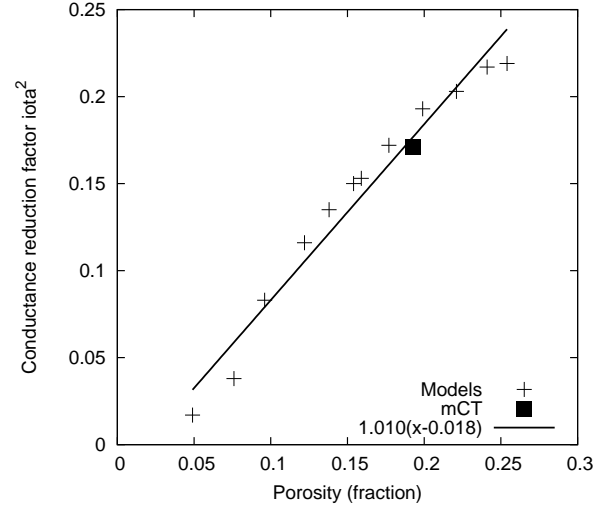}
\caption{Plot showing global conductance reduction factor $\iota^2_g$ versus porosity $\phi$ for the rock models and the micro-CT (mCT) data, together with a linear fit to the data.}
\label{por_vs_iota}
\end{figure}

In Fig. \ref{por_vs_iota} we have plotted the global conductance reduction factor $\iota^2_g$ versus the porosity $\phi$. The linear function $\iota^2_g(\phi)=1.010(\phi-0.018)$ is included as a trend line for the data. For $\phi=0.018$ we have $\iota^2_g(\phi)=0$, which is interpreted to be the percolation threshold for this sandstone \cite{winsauer1952resistivity,mavko1997effect}. Thus 
\begin{equation}
F = \frac{1}{\iota^2_g(\phi) \phi} = \frac{0.989}{\phi(\phi-0.018)} \simeq \frac{1}{\phi(\phi-0.018)}.
\label{eq_por_vs_iota}
\end{equation}
This function is also included in Fig. \ref{por_vs_frf}. The match with the data is fair, given the simplicity of the function.

\begin{figure}
\includegraphics[width=8cm]{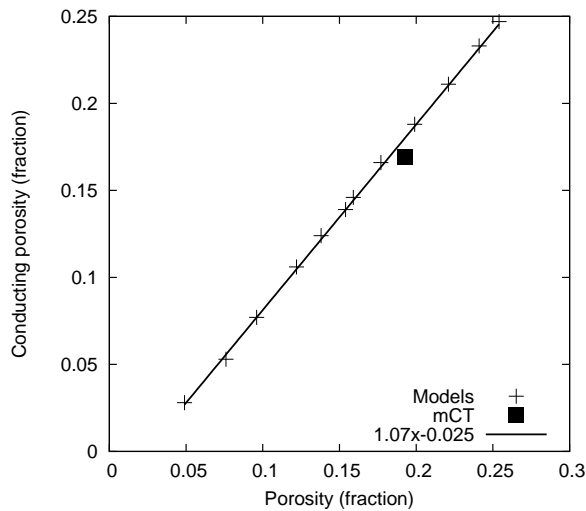}
\caption{Plot showing the correspondence between the porosity $\phi$ and the conducting porosity $\phi_c$ for the rock models and the micro-CT
(mCT) data, together with a linear fit to the data.}
\label{por_vs_gradpor}
\end{figure}

The network volume with nonzero potential gradient can be calculated as
\begin{equation}
\Omega_c = \sum_{\Phi_{t1} \not= \Phi_{t2}}{ \left( \frac{V_{t1}}{\alpha_{t1}} + \frac{V_{t2}}{\alpha_{t2}} + V_{t3} \right) }.
\end{equation}
In Fig. \ref{por_vs_gradpor} we have plotted porosity $\phi$ versus conducting porosity $\phi_c = \Omega_c/V$. A linear fit to the data in Fig. \ref{por_vs_gradpor} gives the correspondence 
\begin{equation}
\phi_c(\phi) = 1.07\phi - 0.025,
\label{phi_path_to_phi}
\end{equation}
which we use in the following. For $\phi=0.025/1.07 = 0.023$ we have $\phi_c(\phi)=0$, which is interpreted as a percolation threshold. This is compared to the threshold $0.018$ derived from the global conductance reduction factor in Eq. \eqref{eq_por_vs_iota}.

For each model and the micro-CT image we discretized the volume $\Omega_c$ into a disjoint union $\sqcup{\Gamma}$, where volume $\Gamma$ conducts a current $I_\Gamma$ such that $\sum{I_\Gamma} = I_t$. Each volume $\Gamma$ is contained in a single series of resistors, with the first resistor connected to an inlet boundary and the last connected to an outlet boundary. Since we cannot trace current across the network nodes in a porous medium, the discretization $\Omega_c = \sqcup{\Gamma}$ is not unique. With respect to a resistor network analog, full mixing is implied at the node points. Different algorithms for the discretization were tested on smaller network analogs for porous media, all giving results comparable to the discretization derived from full mixing at the nodes.

\begin{figure}
\includegraphics[width=8cm]{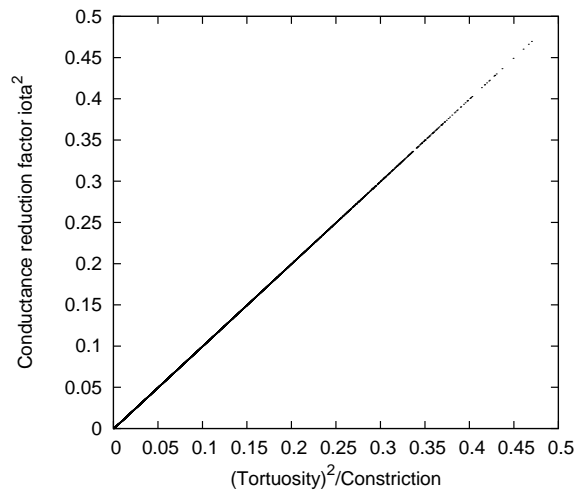}
\caption{Plot showing the correspondence between the conductance reduction factor $\iota(\Gamma)^2$ and the tortuosity squared divided by the constriction $\tau(\Gamma)^2/C(\Gamma)$. Each dot represents a volume $\Gamma$ in the discretization $\Omega_c = \sqcup{\Gamma}$ for the resistor network analog of the micro-CT image of a Bentheimer sandstone.}
\label{iota_vs_tortdivconst}
\end{figure}

The parts $\{ \Gamma_{ti} \}$ of $\Gamma$ is in tubes $i = 1, 2, 3$ of constant cross-sectional area as associated with resistor $t$, as described before. Each volume $\Gamma_{ti} \subset V_{ti}$ then has the associated length $l_{ti}$. For each $\Gamma$ we calculated the conductance reduction factor 
\begin{equation}
\iota^2(\Gamma) = \frac{1}{\Gamma} \sum_t{ \left( \Gamma_{t1} \iota^2_{t1} + \Gamma_{t2} \iota^2_{t2} + \Gamma_{t3} \iota^2_{t3} \right) },
\label{eq:iota_net_path}
\end{equation}
the tortuosity
\begin{equation}
\tau(\Gamma) = \frac{\Delta s}{\sum_t{ l_{t1} +l_{t2} + l_{t3}}},
\end{equation}
and the constriction factor
\begin{align}
C(\Gamma) =& \frac{\Delta \Phi }{\left(\sum_t{l_{t1} +l_{t2} +l_{t3} } \right)^2} \notag \\
& \sum_t{ \left( \frac{l_{t1}^2}{ \lvert \Phi_{t1}-\Phi_{1t} \rvert }+ \frac{l_{t2}^2}{ \lvert \Phi_{t2}-\Phi_{2t} \rvert} + \frac{l_{t3}^2}{ \lvert \Phi_{1t}-\Phi_{2t} \rvert  }\right) }.
\label{eq:const_net_path}
\end{align}
From Fig. \ref{iota_vs_tortdivconst} we see that $\iota(\Gamma)^2$ equals $\tau(\Gamma)^2/C(\Gamma)$ for all $\Gamma$, hence the values from Eqs. \eqref{eq:iota_net_path} to \eqref{eq:const_net_path} are consistent with Eq. \eqref{eq:iota_eq_tortconst_one_path}.

\begin{figure}
\includegraphics[width=8cm]{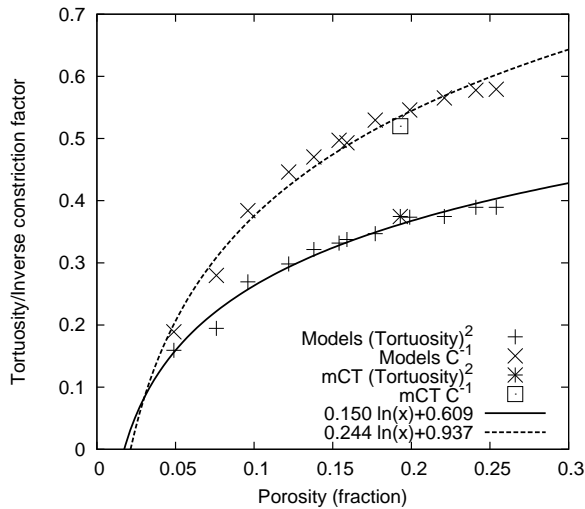}
\caption{Plot showing porosity $\phi$ versus both tortuosity squared $\tau_c^2$ and inverse constriction $C_c^-$ for the rock models and the micro-CT (mCT) data.}
\label{por_vs_tort_const}
\end{figure}

We then obtain the tortuosity and constriction factor for the volume $\Omega_c$:
\begin{align}
\tau^2_c & = \frac{1}{\Omega_c} \sum{\Gamma \tau^2(\Gamma)},
\label{eq:tort_net_total}  \\
C_c & = \frac{1}{I_t} \sum{I_\Gamma C(\Gamma)}.
\label{eq:const_net_total}
\end{align}
The calculated values are listed in Table \ref{tab_models}. We see that $\tau^2_c / C_c = \iota_c^2$, which is consistent with Eq. \eqref{eq:iota_eq_tortconst_global}. The partition of the global conductance reduction factor into tortuosity and constriction gives additional information on the electrical conductance compared to the traditional Archie theory. Note that the values for $\tau_c$ and $\tau$ are similar. 
From Table \ref{tab_models} we observe that $\phi \iota^2_g = \phi_c \iota^2_c = \phi_c \tau^2_c/C_c$, which is in correspondence with the theoretical derivation in Eq. \eqref{eq:iota_tort_const_eq}.

In Fig. \ref{por_vs_tort_const} we have plotted the porosity $\phi$ versus both the tortuosity $\tau_c^2$ and the inverse constriction $C_c^-$, together with functions fitted to the data. For high porosities, i.e., rocks with little cementation, we have less conductivity reduction due to both tortuosity and constriction, compared to that for low porosities. This is as anticipated, as cementation increases the ratio between the pore body and the pore throat cross-sectional area, which yields a larger fluctuation in drift velocity and hence a higher constriction factor. Cementation also can block pore throats completely, which increases the length of the electric field lines and hence reduces the tortuosity. With increasing amounts of cementation (decreasing porosity), tortuosity becomes as important as constriction in reducing the conductivity.

The tortuosity of Bentheimer sandstone is matched fairly well with the function $\tau^2_c(\phi) = 0.150\ln(\phi)+0.609$, and the constriction with the function $C^-_c(\phi) = 0.244\ln(\phi)+0.937$, both plotted in Fig. \ref{por_vs_tort_const}. The function for the constriction indicates a percolation threshold of $e^{-0.937/0.244}=0.020$, while the tortuosity indicates a percolation threshold of $e^{-0.609/0.150}=0.018$, which is in line with the result derived from the global conductance reduction factor in Eq. \eqref{eq_por_vs_iota} and with the result from the conducting porosity in Eq. \eqref{phi_path_to_phi}. Combining Eqs. \eqref{eq:tort_net_total} and \eqref{eq:const_net_total} for tortuosity and constriction with Eq. \eqref{phi_path_to_phi} yields
\begin{align}
&F(\phi) = \frac{1}{\tau^2_c(\phi) C^-_c(\phi) \phi_c} \notag \\
&= \frac{1}{(0.15\ln(\phi)+0.61)(0.24\ln(\phi)+0.94)(1.07\phi-0.025)} \notag \\
&= \frac{26.3}{(\ln(\phi)+4.02)(\ln(\phi)+3.93)(\phi-0.024)} 
\end{align}
This relation is also plotted in Fig. \ref{por_vs_frf} and provides a better match to the Bentheimer data than Eq. \eqref{eq_por_vs_iota}.

\section{Conclusion}

In this work we have described and calculated the electrical conductance of an insulating porous medium filled with an electrolyte. We have shown how the conductance depends on the length of the electric field lines and changes in drift velocity, in addition to the porosity and the conductance of the electrolyte. The lengths of the electric field lines are described by the tortuosity $\tau_c$, and changes in drift velocity are described by the constriction factor $C_c$, both inherent to the microstructure of a porous medium. The combined effect of the tortuosity and constriction is defined as the global conductance reduction factor $\iota^2_c$. When $\phi_c = \phi$, we have $\iota^2_c =\iota^2_g = \tau_c^2/C_c$, where $\iota^2_g$ is the volume weighted average of
$$\iota^2 =  \lVert \nabla \Phi \rVert^2 \left( \frac{\Delta s}{\Delta \Phi} \right)^2.$$

We have shown that our methodology reproduces results for idealized porous media. Our derivation is also applied to natural porous media given by Bentheimer sandstone, where we describe a correlation between porosity and conductivity using the global conductance reduction factor and the distinct contributions calculated for tortuosity $\tau_c$ and constriction $C_c$.

This work enables the substitution of the Archie cementation exponent with a more descriptive, physical parametrization; either with the global conductance reduction factor or with the tortuosity and constriction factors, separately. The latter parameters rely on accurate three-dimensional information on the porous medium microstructure, e.g., micro-CT images. Fortunately, such microstructure data become increasingly more available.

\begin{acknowledgments}
 I would like to thank Rudolf Held (Statoil) for valuable discussions and contributions to the manuscript.
\end{acknowledgments}

\bibliography{myreferences}{}
\bibliographystyle{unsrtnat}

\end{document}